\documentclass[reprint,amsmath,amssymb,aps]{revtex4-1}

\usepackage{graphicx}% Include figure files
\usepackage{dcolumn}% Align table columns on decimal point
\usepackage{bm}% bold math
\usepackage{hyperref}% add hypertext capabilities
\usepackage[mathlines]{lineno}% Enable numbering of text and display math
\usepackage{braket}
\usepackage{lipsum}
\usepackage{alltt}
\usepackage{multirow}

\makeatletter
\allowdisplaybreaks[2]

\newcommand{\e}{{\rm e}}
\newcommand{\tr}{{\mathrm{tr}}}

\newcommand{\dd}{\,\mathrm{d}}
\newcommand{\ra}[1]{\renewcommand{\arraystretch}{#1}}

\newcommand*{\nextone}{%
  \close@column@grid
  \clearpage
  \twocolumngrid}

\begin{document}
\preprint{APS/123-QED}
\title{A Proof of Vivo-Pato-Oshanin's Conjecture \\ on the Fluctuation of von Neumann Entropy}
%\thanks{A footnote to the article title}
\author{Lu Wei}
 \email{luwe@umich.edu}
 \affiliation{Department of Electrical and Computer Engineering \\ University of Michigan-Dearborn, MI 48128, USA}
\date{\today}

\begin{abstract}
It was recently conjectured by Vivo, Pato, and Oshanin~[Phys. Rev. E $\bm{93}$, 052106 (2016)] that for a quantum system of Hilbert dimension $mn$ in a pure state, the variance of the von Neumann entropy of a subsystem of dimension $m\leq n$ is given by
\begin{equation*}
-\psi_{1}\left(mn+1\right)+\frac{m+n}{mn+1}\psi_{1}\left(n\right)-\frac{(m+1)(m+2n+1)}{4n^{2}(mn+1)},
\end{equation*}
where $\psi_{1}(\cdot)$ is the trigamma function. We give a proof of this formula.
\end{abstract}

%\pacs{Valid PACS appear here}% PACS, the Physics and Astronomy Classification Scheme.
%\keywords{Suggested keywords}%Use showkeys class option if keyword display desired

\maketitle

\section{Background and the Conjecture}
Consider a composite quantum system that consists of two subsystems $A$ and $B$ of Hilbert space dimensions $m$ and $n$. The Hilbert space $\mathcal{H}_{A+B}$ of the composite system is given by the tensor product of the Hilbert spaces of the subsystems, $\mathcal{H}_{A+B}=\mathcal{H}_{A}\otimes\mathcal{H}_{B}$. The random pure state of the composite system is written as a linear combination of the random coefficients $x_{i,j}$ and the complete basis $\left\{\Ket{i^{A}}\right\}$ and $\left\{\Ket{j^{B}}\right\}$  of $\mathcal{H}_{A}$ and $\mathcal{H}_{B}$, $\Ket{\psi}=\sum_{i=1}^{m}\sum_{j=1}^{n}x_{i,j}\Ket{i^{A}}\otimes\Ket{j^{B}}$. The corresponding density matrix $\rho=\Ket{\psi}\Bra{\psi}$ has the natural constraint $\tr(\rho)=1$. This implies that the $m\times n$ random coefficient matrix $\mathbf{X}=(x_{i,j})$ satisfies
\begin{equation}\label{eq:pcv}
\tr\left(\mathbf{XX}^{\dag}\right)=1.
\end{equation}
Without loss of generality, it is assumed that $m\leq n$. The reduced density matrix $\rho_{A}$ of the smaller subsystem $A$ admits the Schmidt decomposition $\rho_{A}=\sum_{i=1}^{m}\lambda_{i}\Ket{\phi_{i}^{A}}\Bra{\phi_{i}^{A}}$,
where $\lambda_{i}$ is the $i$-th largest eigenvalue of $\mathbf{XX}^{\dag}$. The conservation of probability~(\ref{eq:pcv}) now implies the constraint $\sum_{i=1}^{m}\lambda_{i}=1.$ The probability measure of the random coefficient matrix $\mathbf{X}$ is the Haar measure, where the entries are uniformly distributed over all the possible values satisfying the constraint~(\ref{eq:pcv}). The resulting eigenvalue density of $\mathbf{XX}^{\dag}$ is well known (see, e.g.,~\cite{Page1993}),
\begin{eqnarray}
f\left(\bm{\lambda}\right)&=&\frac{\Gamma(mn)}{c}~\delta\left(1-\sum_{i=1}^{m}\lambda_{i}\right)\times\nonumber\label{eq:fte} \\
&&\prod_{1\leq i<j\leq m}\left(\lambda_{i}-\lambda_{j}\right)^{2}\prod_{i=1}^{m}\lambda_{i}^{n-m},
\end{eqnarray}\\
where $\delta(\cdot)$ is the Dirac delta function and the constant
\begin{equation}\label{eq:con}
c=\prod_{i=1}^{m}\Gamma(n-i+1)\Gamma(i).
\end{equation}
The random matrix ensemble~(\ref{eq:fte}) is also known as the (unitary) fixed-trace ensemble. The considered bipartite quantum system is a fundamental model that describes the interaction between physical object and its environment. For example~\cite{Page1993}, the subsystem $A$ is the black hole and the subsystem $B$ is the associated radiation field. In another example~\cite{Majumdar}, the subsystem $A$ is a set of spins and the subsystem $B$ represents the environment of a heat bath.

A measure of the entanglement of the considered bipartite quantum system is the von Neumann entropy
\begin{equation}\label{eq:vN}
S=-\sum_{i=1}^{m}\lambda_{i}\ln\lambda_{i},
\end{equation}
where $S\in\left[0, \ln{m}\right]$. Its mean value was conjectured by Page~\cite{Page1993} as
\begin{equation}\label{eq:vNm}
\mathbb{E}_{f}\!\left[S\right]=\psi_{0}(mn+1)-\psi_{0}(n)-\frac{m+1}{2n},
\end{equation}
where $\mathbb{E}_{f}\!\left[\cdot\right]$ denotes that the expectation is taken over the fixed-trace ensemble~(\ref{eq:fte}). Here, $\psi_{0}(x)=\dd\ln\Gamma(x)/\dd x$ is the digamma function (Psi function)~\cite{Luke} and for a positive integer $l$,
\begin{equation}\label{eq:dg}
\psi_{0}(l)=-\gamma+\sum_{k=1}^{l-1}\frac{1}{k},
\end{equation}
where $\gamma\approx0.5772$ is the Euler's constant. The mean value formula~(\ref{eq:vNm}) was proved independently by Foong-Kanno~\cite{Foong1994}, S\'{a}nchez-Ruiz~\cite{Ruiz1995}, Sen~\cite{Sen1996}, and Adachi-Toda-Kubotani~\cite{Adachi2009}. For the orthogonal and symplectic fixed-trace ensembles, the mean formulas of the von Neumann entropy were derived by Kumar-Pandey~\cite{Kumar2011}.

To gain more insights, one needs to know the fluctuation of the von Neumann entropy. In fact, its mean value turns out to be a poor representative that has led to an incorrect conclusion on the full distribution~\cite{Page1993}. Recently, Vivo, Pato, and Oshanin conjectured~\cite[eq.~(57)]{VPO16}, based on small $n$ and $m$ calculations from some complicated representations~\cite[eqs.~(54)--(56),~(A3),~(A9)]{VPO16}, that the variance of the von Neumann entropy $\mathbb{V}\!_{f}\!\left[S\right]$ equals
\begin{equation}\label{eq:vNv}
-\psi_{1}\left(mn+1\right)+\frac{m+n}{mn+1}\psi_{1}\left(n\right)-\frac{(m+1)(m+2n+1)}{4n^{2}(mn+1)},
\end{equation}
where $\psi_{1}(x)=\dd^{2}\ln\Gamma(x)/\dd x^{2}$ is the trigamma function~\cite{Luke}\footnote{The digamma and trigamma functions are the polygamma functions of order zero and one, respectively.} and for a positive integer $l$,
\begin{equation}\label{eq:tg}
\psi_{1}(l)=\frac{\pi^{2}}{6}-\sum_{k=1}^{l-1}\frac{1}{k^{2}}.
\end{equation}

In this paper, we show that the conjecture~(\ref{eq:vNv}) of Vivo-Pato-Oshanin (VPO) is indeed correct. The presentation of the proof is organized as follows. In Sec.~\ref{sec:relation}, we relate the variance of the von Neumann entropy to that of an induced one over the Laguerre ensemble, which is calculated explicitly. The derived induced variance is simplified to functions involving digamma and trigamma functions in Sec.~\ref{sec:clear} that leads to a proof of the conjecture. Most of the technical tools for the simplification are presented in the Appendix. Finally, we point out that even though the exact distribution of von Neumann entropy is unknown, its asymptotic distribution was obtained via the Coulomb gas approach by Nadal-Majumdar-Vergassola~\cite{Nadal2011}.

\section{Variance of an Induced Entropy in Laguerre Ensemble}\label{sec:relation}
\subsection{Variance Relation}
By the construction~(\ref{eq:pcv}), the random coefficient matrix $\mathbf{X}$ has a natural relation with a Wishart matrix $\mathbf{YY}^{\dag}$ as
\begin{equation}\label{eq:wf}
\mathbf{XX}^{\dag}=\frac{\mathbf{YY}^{\dag}}{\tr\left(\mathbf{YY}^{\dag}\right)},
\end{equation}
where $\mathbf{Y}$ is an $m\times n$ ($m\leq n$) matrix of independently and identically distributed complex Gaussian entries. The density of the eigenvalues $0<\theta_{m}<\dots<\theta_{1}<\infty$ of $\mathbf{YY}^{\dag}$ equals~\cite{Forrester}
\begin{equation}\label{eq:we}
g\left(\bm{\theta}\right)=\frac{1}{c}\prod_{1\leq i<j\leq m}\left(\theta_{i}-\theta_{j}\right)^{2}\prod_{i=1}^{m}\theta_{i}^{n-m}\e^{-\theta_{i}},
\end{equation}
where $c$ is given by~(\ref{eq:con}) and the above ensemble is known as the Laguerre ensemble. The trace of the Wishart matrix
\begin{equation}
r=\tr\left(\mathbf{YY}^{\dag}\right)=\sum_{i=1}^{m}\theta_{i}
\end{equation}
follows a gamma distribution with the density~\cite{VPO16}
\begin{equation}\label{eq:r}
h_{mn}(r)=\frac{1}{\Gamma(mn)}\e^{-r}r^{mn-1},~~~r\in[0,\infty).
\end{equation}
The relation~(\ref{eq:wf}) induces the change of variables
\begin{equation}\label{eq:cv}
\lambda_{i}=\frac{\theta_{i}}{r},~~~i=1,\ldots,m,
\end{equation}
that leads to a well-known relation (see, e.g.~\cite{Page1993}) among the densities~(\ref{eq:fte}),~(\ref{eq:we}), and~(\ref{eq:r}) as
\begin{equation}\label{eq:relation}
f\left(\bm{\lambda}\right)h_{mn}(r)\dd r\prod_{i=1}^{m}\dd\lambda_{i}=g\left(\bm{\theta}\right)\prod_{i=1}^{m}\dd\theta_{i}.
\end{equation}
This relation implies that $r$ is independent of each $\lambda_{i}$, $i=1,\ldots,m$, since their densities factorize.

Page~\cite{Page1993} exploited the relation~(\ref{eq:relation}) by relating the first moment of von Neumann entropy over the fixed-trace ensemble~(\ref{eq:fte}) to that of an induced entropy~\footnote{For convenience of the discussion, we refer the random variable $T$ as an induced entropy, which may not have physical meaning of an entropy.}
\begin{equation}\label{eq:ivN}
T=\sum_{i=1}^{m}\theta_{i}\ln\theta_{i},
\end{equation}
over the Laguerre ensemble~(\ref{eq:we}) as follows. First, by using the relations~(\ref{eq:cv}), one has
\begin{equation}\label{eq:ST}
S=-\sum_{i=1}^{m}\frac{\theta_{i}}{r}\ln\frac{\theta_{i}}{r}=r^{-1}\left(r\ln r-T\right).
\end{equation}
Then, the expected value of $S$ is evaluated as
\begin{eqnarray}
\mathbb{E}_{f}\!\left[S\right]&=&\int_{\bm{\lambda}}r^{-1}\left(r\ln r-T\right)f\left(\bm{\lambda}\right)\prod_{i=1}^{m}\dd\lambda_{i}\nonumber\\
&=&\int_{\bm{\lambda}}r^{-1}\left(r\ln r-T\right)f\left(\bm{\lambda}\right)\prod_{i=1}^{m}\dd\lambda_{i}\int_{r}h_{mn+1}(r)\dd r\nonumber\\
&=&\frac{1}{mn}\int_{\bm{\lambda}}f\left(\bm{\lambda}\right)\prod_{i=1}^{m}\dd\lambda_{i}\int_{r}h_{mn}(r)r\ln r\dd r-\nonumber\\
&&\frac{1}{mn}\int_{\bm{\lambda}}\int_{r}T f\left(\bm{\lambda}\right)h_{mn}(r)\dd r\prod_{i=1}^{m}\dd\lambda_{i}\label{eq:ES1}\\
&=&\psi_{0}(mn+1)-\frac{1}{mn}\mathbb{E}_{g}\!\left[T\right],\label{eq:ES2}
\end{eqnarray}
where the expectation $\mathbb{E}_{g}\!\left[\cdot\right]$ is taken over the Laguerre ensemble~(\ref{eq:we}). Here, (\ref{eq:ES1}) is obtained by the identity $r^{-1}h_{mn+1}(r)=h_{mn}(r)/mn$ and the fact that $r$ is independent of $\bm{\lambda}$, and~(\ref{eq:ES2}) is established by the change of measures~(\ref{eq:relation}) and the identity
\begin{equation}\label{eq:1eln}
\int_{0}^{\infty}\!\!\e^{-r}r^{a-1}\ln{r}\dd r=\Gamma(a)\psi_{0}(a),~~~\operatorname{Re}(a)>0.
\end{equation}
S\'{a}nchez-Ruiz~\cite{Ruiz1995} and Sen~\cite{Sen1996} have calculated that
\begin{equation}\label{eq:Tm}
\mathbb{E}_{g}\!\left[T\right]=mn\psi_{0}(n)+\frac{1}{2}m(m+1),
\end{equation}
and together with the relation~(\ref{eq:ES2}) leads to their proofs of Page's conjecture on the mean entropy~(\ref{eq:vNm}).

We now show that the idea of Page~\cite{Page1993} can be generalized to find a relation between the second moments (hence the variances since the first moments are known) of $S$ and $T$, which is the starting point of our calculations. First, using the result~(\ref{eq:ST}) we have
\begin{eqnarray}
S^{2}&=&r^{-2}\left(T^{2}-T~2r\ln{r}+r^{2}\ln^{2}{r}\right)\label{eq:ST1}\\
&=&r^{-2}\left(T^{2}+S~2r^{2}\ln{r}-r^{2}\ln^{2}{r}\right).\label{eq:ST2}
\end{eqnarray}
The expression~(\ref{eq:ST2}) is obtained by replacing only the first power of $T$ in~(\ref{eq:ST1}) by the identity~(\ref{eq:ST}), and the reason for this replacement will become clear. The second moment of $S$ can now be written as
\begin{equation}\label{eq:VS}
\mathbb{E}_{f}\!\left[S^{2}\right]=\int_{\bm{\lambda}}r^{-2}\left(T^{2}+S~2r^{2}\ln{r}-r^{2}\ln^{2}{r}\right)f\left(\bm{\lambda}\right)\prod_{i=1}^{m}\dd\lambda_{i}.
\end{equation}
To utilize the independence between $r$ and $\bm{\lambda}$, we multiple~(\ref{eq:VS}) by an appropriate constant $1=\int_{r}h_{mn+2}(r)\dd{r}$, which, with the fact that $r^{-2}h_{mn+2}(r)=h_{mn}(r)/mn(mn+1)$, leads to
\begin{eqnarray}
&&\mathbb{E}_{f}\!\left[S^{2}\right]=\frac{1}{mn(mn+1)}\int_{\bm{\lambda}}\int_{r}T^{2}f\left(\bm{\lambda}\right)h_{mn}(r)\dd r\prod_{i=1}^{m}\dd\lambda_{i}+\nonumber\\
&&\frac{2}{mn(mn+1)}\int_{\bm{\lambda}}Sf\left(\bm{\lambda}\right)\prod_{i=1}^{m}\dd\lambda_{i}\int_{r}h_{mn}(r)r^{2}\ln{r}\dd{r}-\nonumber\\
&&\frac{1}{mn(mn+1)}\int_{\bm{\lambda}}f\left(\bm{\lambda}\right)\prod_{i=1}^{m}\dd\lambda_{i}\int_{r}h_{mn}(r)r^{2}\ln^{2}{r}\dd{r}.\nonumber
\end{eqnarray}
From the second line of the above equation, we see that the replacement of the first power of $T$ by $S$ in~(\ref{eq:ST1}) makes it possible to evaluate the integrals over $r$ and $\bm{\lambda}$ separately. Finally, using the change of measures~(\ref{eq:relation}) as well as the identities~(\ref{eq:1eln}) and
\begin{equation}\label{eq:2eln}
\int_{0}^{\infty}\!\!\e^{-r}r^{a-1}\ln^{2}{r}\dd{r}=\Gamma(a)\left(\psi_{1}(a)+\psi_{0}^{2}(a)\right),~\operatorname{Re}(a)>0,
\end{equation}
we arrive at
\begin{eqnarray}
\mathbb{E}_{f}\!\left[S^{2}\right]&=&\frac{1}{mn(mn+1)}\mathbb{E}_{g}\!\left[T^{2}\right]+2\psi_{0}(mn+2)\mathbb{E}_{f}\!\left[S\right]-\nonumber\\
&&\psi_{1}(mn+2)-\psi_{0}^{2}(mn+2).\label{eq:VS1}
\end{eqnarray}
Inserting the mean formula~(\ref{eq:vNm}) and the VPO's conjecture~(\ref{eq:vNv}) into the definition $\mathbb{E}_{f}\!\left[S^2\right]=\mathbb{V}\!_{f}\!\left[S\right]+\mathbb{E}_{f}^{2}\!\left[S\right]$, and equating it to the derived relation~(\ref{eq:VS1}), the VPO's conjecture boils down to showing that $\mathbb{E}_{g}\!\left[T^{2}\right]$ is given by
\begin{widetext}
\begin{equation}\label{eq:eqc}
mn(m+n)\psi_{1}(n)+mn(mn+1)\psi_{0}^{2}(n)+m\left(m^{2}n+mn+m+2n+1\right)\psi_{0}(n)+\frac{1}{4}m(m+1)\left(m^2+m+2\right),
\end{equation}
\end{widetext}
where we have used the identities (cf.~(\ref{eq:dg}) and~(\ref{eq:tg}))
\begin{subequations}\label{eq:pgre}
\begin{eqnarray}
\psi_{0}(l+n)&=&\psi_{0}(l)+\sum_{k=0}^{n-1}\frac{1}{l+k},\label{eq:dgre}\\
\psi_{1}(l+n)&=&\psi_{1}(l)-\sum_{k=0}^{n-1}\frac{1}{(l+k)^2},\label{eq:tgre}
\end{eqnarray}
\end{subequations}
for the case $l=mn+1$, $n=1$.

We have so far converted the VPO's conjecture~(\ref{eq:vNv}) evaluated over the fixed-trace ensemble~(\ref{eq:fte}) to an equivalent conjecture~(\ref{eq:eqc}) evaluated over the Laguerre ensemble~(\ref{eq:we}). Instead of working directly with the complicated correlation functions of the fixed-trace ensemble as in~\cite{Adachi2009,Kumar2011,VPO16}, the induced variance over the well-investigated correlation functions of the Laguerre ensemble can be explicitly calculated as will be shown in Sec.~\ref{sec:cal}. The proposed `moments conversion' approach may be generalized to study the higher moments of the von Neumann entropy as well as other entanglement measures such as the Tsallis entropy and the R\'{e}nyi entropy.

\subsection{Calculations of the Induced Variance}\label{sec:cal}
Since $T^{2}=\sum_{i=1}^{m}\theta_{i}^{2}\ln^{2}\theta_{i}+2\sum_{1\leq i<j\leq m}\theta_{i}\theta_{j}\ln\theta_{i}\ln\theta_{j}$, the calculation of $\mathbb{E}_{g}\!\left[T^{2}\right]$ involves one and two arbitrary eigenvalue densities, denoted respectively by $g_{1}(x_{1})$ and $g_{2}(x_{1},x_{2})$, of the Laguerre ensemble as
\begin{eqnarray}\label{eq:i2m}
\mathbb{E}_{g}\!\left[T^{2}\right]&=&m\int_{0}^{\infty}\!\!x_{1}^{2}\ln^{2}x_{1}~g_{1}(x_{1})\dd x_{1}+m(m-1)\times\nonumber\\
\int_{0}^{\infty}\!\!\!\!\!\!\!\!\!&&\int_{0}^{\infty}\!\!x_{1}x_{2}\ln{x_{1}}\ln{x_{2}}~g_{2}\left(x_{1},x_{2}\right)\dd x_{1}\dd x_{2}.
\end{eqnarray}
In general, the joint density of $N$ arbitrary eigenvalues $g_{N}(x_{1},\dots,x_{N})$ is related to the $N$-point correlation function
\begin{equation}\label{eq:cf}
X_{N}\left(x_{1},\dots,x_{N}\right)=\det\left(K\left(x_{i},x_{j}\right)\right)_{i,j=1}^{N}
\end{equation}
as~\cite{Forrester} $g_{N}(x_{1},\dots,x_{N})=X_{N}\left(x_{1},\dots,x_{N}\right)(m-N)!/m!$, where $\det(\cdot)$ is the matrix determinant and the symmetric function $K(x_{i},x_{j})$ is the correlation kernel. In particular, we have
\begin{eqnarray*}
g_{1}(x_{1})&=&\frac{1}{m}K(x_{1},x_{1}),\\
g_{2}(x_{1},x_{2})&=&\frac{1}{m(m-1)}\left(K(x_{1},x_{1})K(x_{2},x_{2})-K^{2}(x_{1},x_{2})\right).
\end{eqnarray*}
As a result, one can represent~(\ref{eq:i2m}) as
\begin{equation}\label{eq:i2m1}
\mathbb{E}_{g}\!\left[T^{2}\right]=I_{A}-I_{B}+\left(mn\psi_{0}(n)+\frac{1}{2}m(m+1)\right)^2,
\end{equation}
where
\begin{eqnarray}
\!\!\!\!\!\!\!\!\!I_{A}&=&\int_{0}^{\infty}\!\!x_{1}^{2}\ln^{2}x_{1}~K(x_{1},x_{1})\dd x_{1},\label{eq:IA}\\
\!\!\!\!\!\!\!\!\!I_{B}&=&\int_{0}^{\infty}\!\!\int_{0}^{\infty}\!\!x_{1}x_{2}\ln{x_{1}}\ln{x_{2}}~K^{2}\left(x_{1},x_{2}\right)\dd x_{1}\dd x_{2},\label{eq:IB}
\end{eqnarray}
and we have used the result~(\ref{eq:Tm}) and the definition
\begin{equation*}
\int_{0}^{\infty}\!\!x\ln{x}~K(x,x)\dd{x}=m\int_{0}^{\infty}\!\!x\ln{x}~g_{1}(x)\dd{x}=\mathbb{E}_{g}\!\left[T\right].
\end{equation*}

Before computing the integrals $I_{A}$ and $I_{B}$, the following results on the correlation functions~(\ref{eq:cf}) are needed. The correlation kernel of the Laguerre ensemble can be explicitly written as~\cite{Forrester}
\begin{equation}\label{eq:ker}
K(x_{i},x_{j})=\sqrt{\e^{-x_{i}-x_{j}}(x_{i}x_{j})^{n-m}}\sum_{k=0}^{m-1}\frac{C_{k}(x_{i})C_{k}(x_{j})}{k!(n-m+k)!},
\end{equation}
where
\begin{equation}
C_{k}(x)=(-1)^{k}k!L_{k}^{(n-m)}(x)
\end{equation}
with
\begin{equation}
L_{k}^{(n-m)}(x)=\sum_{i=0}^{k}(-1)^{i}\binom{n-m+k}{k-i}\frac{x^i}{i!}
\end{equation}
being the (generalized) Laguerre polynomial of degree $k$. The Laguerre polynomials satisfy the well-known orthogonality relation~\cite{Forrester}
\begin{equation}\label{eq:oc}
\int_{0}^{\infty}\!\!x^{n-m}\e^{-x}L_{k}^{(n-m)}(x)L_{l}^{(n-m)}(x)\dd{x}=\frac{(n-m+k)!}{k!}\delta_{kl},
\end{equation}
where $\delta_{kl}$ is the Kronecker delta function. It is known that the one-point correlation function (cf.~(\ref{eq:cf})) admits a more convenient representation as~\cite{Ruiz1995,Forrester}
\begin{eqnarray}\label{eq:one}
X_{1}(x)&=&\frac{m!}{(n-1)!}x^{n-m}\e^{-x}\bigg(\left(L_{m-1}^{(n-m+1)}(x)\right)^{2}-\nonumber\\
&&L_{m-2}^{(n-m+1)}(x)L_{m}^{(n-m+1)}(x)\bigg).
\end{eqnarray}
We also need the following identity, due to Schr{\"{o}}dinger~\cite{Schrodinger1926}, that generalizes the integral~(\ref{eq:oc}) to
\begin{eqnarray}\label{eq:Swm}
\!\!\!\!\!\!\!\!&&\int_{0}^{\infty}\!\!x^{q}\e^{-x}L_{s}^{(\alpha)}(x)L_{t}^{(\beta)}(x)\dd{x}=(-1)^{s+t}\times\nonumber \\
\!\!\!\!\!\!\!\!&&\sum_{k=0}^{\min(s,t)}\binom{q-\alpha}{s-k}\binom{q-\beta}{t-k}\frac{\Gamma(q+1+k)}{k!},~~~q>-1.
\end{eqnarray}
By taking the first and second derivative on both sides of~(\ref{eq:Swm}) with respect to $q$, we obtain two more integral identities as shown in~(\ref{eq:1d}) (see also~\cite{Ruiz1995}) and~(\ref{eq:2d}), which are respectively denoted by $B_{s,t}^{(\alpha,\beta)}(q)$ and $A_{s,t}^{(\alpha,\beta)}(q)$. With the above preparations, we now proceed to the calculations of $I_{A}$ in~(\ref{eq:IA}) and $I_{B}$ in~(\ref{eq:IB}).

\subsubsection{Calculating $I_{A}$}\label{sec:IA}
By the fact that~(cf.~(\ref{eq:cf}))
\begin{equation}
X_{1}(x_{1})=K(x_{1},x_{1}),
\end{equation}
one inserts~(\ref{eq:one}) into~(\ref{eq:IA}) to obtain
\begin{equation}\label{eq:IAi}
I_{A}=\frac{m!}{(n-1)!}\left(\mathcal{A}_{m-1,m-1}-\mathcal{A}_{m-2,m}\right),
\end{equation}
where for convenience we have further defined (cf.~(\ref{eq:2d}))
\begin{equation}
\mathcal{A}_{s,t}=A_{s,t}^{(n-m+1,n-m+1)}(n-m+2).
\end{equation}
We now use~(\ref{eq:2d}), and the contribution to the sum
\begin{eqnarray}
\!\!\!&&\mathcal{A}_{m-1,m-1}=\sum_{k=0}^{m-1}\binom{1}{m-1-k}^{2}\frac{(n-m+2+k)!}{k!}\times\nonumber\\
\!\!\!&&\big(\!\left(\psi_{0}(n-m+3+k)+2\psi_{0}(2)-2\psi_{0}(3-m+k)\right)^{2}+\nonumber\\
\!\!\!&&\psi_{1}(n-m+3+k)+2\psi_{1}(2)-2\psi_{1}(3-m+k)\big)
\end{eqnarray}\\
consists of the cases when the binomial terms are zero ($k=0,\dots,m-3$) with the polygamma functions being infinity and are nonzero ($k=m-2, m-1$) with the polygamma functions being finite. Namely, we have
\begin{multline}\label{eq:IA0}
\mathcal{A}_{m-1,m-1}=\frac{(n+1)!}{(m-1)!}\left(\psi_{0}^{2}(n+2)+\psi_{1}(n+2)\right)+\\
\frac{n!}{(m-2)!}\left((\psi_{0}(n+1)+2)^{2}+\psi_{1}(n+1)-2\right)+\\
\sum_{k=0}^{m-3}\frac{(n-m+2+k)!}{(m-1-k)!^{2}k!}\frac{4\psi_{0}^{2}(3-m+k)-2\psi_{1}(3-m+k)}{\Gamma^{2}(3-m+k)},
\end{multline}
% \nextone\noindent
\begin{widetext}
\begin{eqnarray}\label{eq:1d}
&&B_{s,t}^{(\alpha,\beta)}(q)=\int_{0}^{\infty}\!\!x^{q}\e^{-x}\ln{x}~L_{s}^{(\alpha)}(x)L_{t}^{(\beta)}(x)\dd{x}=(-1)^{s+t}\sum_{k=0}^{\min(s,t)}\binom{q-\alpha}{s-k}\binom{q-\beta}{t-k}\frac{\Gamma(q+1+k)}{k!}\times\nonumber\\
&&\big(\psi_{0}(q+1+k)+\psi_{0}(q-\alpha+1)+\psi_{0}(q-\beta+1)-\psi_{0}(q-\alpha-s+1+k)-\psi_{0}(q-\beta-t+1+k)\big).
\end{eqnarray}
\hrulefill
\begin{eqnarray}\label{eq:2d}
&&A_{s,t}^{(\alpha,\beta)}(q)=\int_{0}^{\infty}\!\!x^{q}\e^{-x}\ln^{2}{x}~L_{s}^{(\alpha)}(x)L_{t}^{(\beta)}(x)\dd{x}=(-1)^{s+t}\sum_{k=0}^{\min(s,t)}\binom{q-\alpha}{s-k}\binom{q-\beta}{t-k}\frac{\Gamma(q+1+k)}{k!}\times\nonumber\\
&&\Big(\big(\psi_{0}(q+1+k)+\psi_{0}(q-\alpha+1)+\psi_{0}(q-\beta+1)-\psi_{0}(q-\alpha-s+1+k)-\psi_{0}(q-\beta-t+1+k)\big)^{2}+\nonumber\\
&&\psi_{1}(q+1+k)+\psi_{1}(q-\alpha+1)+\psi_{1}(q-\beta+1)-\psi_{1}(q-\alpha-s+1+k)-\psi_{1}(q-\beta-t+1+k)\Big).
\end{eqnarray}
\end{widetext}
which by interpreting the gamma and polygamma functions of negative integer arguments as the limit $\epsilon\to0$ of
\begin{subequations}\label{eq:pgna}
\begin{eqnarray}
\!\!\!\!\!\!\!\!\!\!\!\!\Gamma(-l+\epsilon)&=&\frac{(-1)^{l}}{l!\epsilon}\left(1+\psi_{0}(l+1)\epsilon+o\left(\epsilon^2\right)\right),\label{eq:pgna1}\\
\!\!\!\!\!\!\!\!\!\!\!\!\psi_{0}(-l+\epsilon)&=&-\frac{1}{\epsilon}\left(1-\psi_{0}(l+1)\epsilon+o\left(\epsilon^2\right)\right),\label{eq:pgna2}\\
\!\!\!\!\!\!\!\!\!\!\!\!\psi_{1}(-l+\epsilon)&=&\frac{1}{\epsilon^2}\left(1+o\left(\epsilon^2\right)\right),\label{eq:pgna3}
\end{eqnarray}
\end{subequations}
leads to a well-defined limit
\begin{eqnarray}\label{eq:IA1}
&&\frac{1}{(m-1-k)!^{2}}\frac{4\psi_{0}^{2}(3-m+k)-2\psi_{1}(3-m+k)}{\Gamma^{2}(3-m+k)}=\nonumber\\
&&\frac{2}{(m-2-k)^{2}(m-1-k)^{2}},~~~k=0,\ldots,m-3.
\end{eqnarray}
In the same manner that has led to $\mathcal{A}_{m-1,m-1}$, we obtain
\begin{eqnarray}\label{eq:IA2}
&&\mathcal{A}_{m-2,m}=\frac{n!}{(m-2)!}\left(\psi_{0}(n+1)+1\right)-\nonumber\\
&&\frac{(n-1)!}{3(m-3)!}\left(\psi_{0}(n)+1\right)+\sum_{k=0}^{m-4}\frac{(n-m+2+k)!}{k!}\times\nonumber\\
&&\frac{2}{(m-3-k)(m-2-k)(m-1-k)(m-k)}.
\end{eqnarray}

Finally, we insert~(\ref{eq:IA0}),~(\ref{eq:IA1}),~(\ref{eq:IA2}) into~(\ref{eq:IAi}) and simplify the expression by rearranging the sums as well as using~(\ref{eq:pgre}) to obtain
\begin{widetext}
\begin{eqnarray}\label{eq:IAe}
I_{A}&=&\frac{m}{3}\left(3n(m+n)\psi_{1}(n)+3n(m+n)\psi_{0}^{2}(n)+(m^{2}+9mn+3m+3n+2)\psi_{0}(n)+m^{2}+3mn+6m-3n-1\right)+\nonumber\\
&&\frac{2m!}{(n-1)!}\left(\sum_{k=1}^{m-2}\frac{(n-k)!}{(m-2-k)!}\frac{1}{k^{2}(k+1)^{2}}-\sum_{k=1}^{m-3}\frac{(n-1-k)!}{(m-3-k)!}\frac{1}{k(k+1)(k+2)(k+3)}\right).
\end{eqnarray}
\end{widetext}

\subsubsection{Calculating $I_{B}$}\label{sec:IB}
Inserting~(\ref{eq:ker}) into~(\ref{eq:IB}) and using the symmetry of the correlation kernel, the integral $I_{B}$ can be represented as
\begin{eqnarray}\label{eq:IBi}
I_{B}&=&2\sum_{j=1}^{m-1}\sum_{k=0}^{m-j-1}\frac{k!(k+j)!\mathcal{B}_{k+j,k}^{2}}{(n-m+k)!(n-m+k+j)!}+\nonumber\\ &&\sum_{k=0}^{m-1}\frac{k!^2\mathcal{B}_{k,k}^{2}}{(n-m+k)!^2},
\end{eqnarray}
where we have further defined (cf.~(\ref{eq:1d}))
\begin{equation}
\mathcal{B}_{s,t}=B_{s,t}^{(n-m,n-m)}(n-m+1).
\end{equation}
The identity~(\ref{eq:1d}) gives
\begin{eqnarray}\label{eq:IB1}
\mathcal{B}_{k,k}&=&\sum_{j=0}^{k}\binom{1}{k-j}^{2}\frac{(n-m+1+j)!}{j!}\big(2\psi_{0}(2)+\nonumber\\
&&\psi_{0}(n-m+2+j)-2\psi_{0}(2-k+j)\big)\nonumber\\
&=&\frac{(n-m+k)!}{k!}\big((n-m+1+2k)\times\nonumber\\
&&\psi_{0}(n-m+1+k)+2k+1\big),
\end{eqnarray}
where $j=k-1,k$ provides the nonzero contribution to the sum and we have used~(\ref{eq:dgre}) for the simplification. In the same manner, one obtains
\begin{eqnarray}\label{eq:IB2}
\!\!\!\!\!\!\!\!\!\!\!\!\mathcal{B}_{k+1,k}&=&\frac{(n-m+k)!}{k!}\big((n-m+1+k)\times\nonumber\\
\!\!\!\!\!\!\!\!\!\!\!\!&&\psi_{0}(n-m+1+k)+n-m+3k/2+2\big)
\end{eqnarray}\\
and the cases $j=2,\dots,m-1$ are computed to be
\begin{equation}\label{eq:IB3}
\mathcal{B}_{k+j,k}=\frac{(n-m+k)!}{k!j}\left(\frac{n-m+1+k}{j-1}-\frac{k}{j+1}\right).
\end{equation}

Inserting~(\ref{eq:IB1}),~(\ref{eq:IB2}), and~(\ref{eq:IB3}) into~(\ref{eq:IBi}), we arrive at
\begin{widetext}
\begin{eqnarray}\label{eq:IBe}
I_{B}&=&\sum_{k=0}^{m-1}\left(\left(n-m+1+2k\right)\psi_{0}(n-m+1+k)+2k+1\right)^{2}+\sum_{k=0}^{m-2}\frac{2(k+1)}{n-m+1+k}((n-m+1+k)\psi_{0}(n-m+1+k)+\nonumber\\
&&n-m+2+3k/2)^{2}+\sum_{j=2}^{m-1}\sum_{k=0}^{m-1-j}\frac{2(n-m+k)!(k+j)!}{(n-m+k+j)!k!j^{2}}\left(\frac{n-m+1+k}{j-1}-\frac{k}{j+1}\right)^{2}.
\end{eqnarray}
\end{widetext}

\section{Simplification of Summations}\label{sec:clear}
The remaining task is to simplify the sums appear in $I_{A}$ and $I_{B}$ to polygamma functions. This is a straightforward but tedious task, for which we need several finite sum identities as listed in the Appendix. Some remarks on these identities are also provided in the Appendix. Though $I_{A}$ in~(\ref{eq:IAe}) and $I_{B}$ in~(\ref{eq:IBe}) are valid for any positive integers $m$ and $n$ with $m\leq n$, as will be seen it is convenient to assume $n>m\geq3$ in the following simplification. For this reason, we will first simplify $I_{A}$ and $I_{B}$ in the case $n>m\geq3$. The remaining special cases will be considered at the end of this section.

For ease of presentation, we cite the identities used in each step on top of the equality symbol. The argument of each of the resulting polygamma functions is shifted to one of the following $n-m+2$, $m$, $n$, $1$, with the help of~(\ref{eq:pgre}). In addition, simplification by combining like terms is also performed in each step without being explicitly mentioned. We start with $I_{A}$ in~(\ref{eq:IAe}), where by using partial fraction decomposition the first sum is simplified as
\begin{widetext}
\begin{eqnarray}\label{eq:IAs1}
&&\sum_{k=1}^{m-2}\frac{(n-k)!}{(m-2-k)!}\frac{1}{k^{2}(k+1)^{2}}=\sum_{k=1}^{m-2}\frac{(n-k)!}{(m-2-k)!}\left(\frac{1}{k^2}+\frac{1}{(k+1)^2}+\frac{2}{k+1}-\frac{2}{k}\right)\nonumber\\
&=&(m+n)\sum_{k=2}^{m-2}\frac{(n-k)!}{(m-1-k)!}\frac{1}{k^{2}}+2(n-m+1)\sum_{k=1}^{m-1}\frac{(n-k)!}{(m-1-k)!}\frac{1}{k}+\frac{(m+n)(n-m+1)!}{(m-1)^2}-\frac{(2n-m)(n-1)!}{(m-2)!}\nonumber\\
&\stackrel{(\ref{eq:sum11})}{=}&(m+n)\sum_{k=1}^{m-1}\frac{(n-k)!}{(m-1-k)!}\frac{1}{k^{2}}+\frac{2(n-m+1)n!}{(m-1)!}(\psi_{0}(n)-\psi_{0}(n-m+2))+\frac{(5n-3mn-2m+2)(n-1)!}{(m-1)!}\nonumber\\
&\stackrel{(\ref{eq:sum12})}{=}&\frac{(m+n)n!}{(m-1)!}\sum_{k=1}^{m}\frac{\psi_{0}(n-m+k)}{k}+\frac{(m+n)n!}{(m-1)!}\bigg(\frac{1}{2}\psi_{1}(n-m+2)-\frac{1}{2}\psi_{1}(n)-\frac{1}{2}\psi_{0}^{2}(n-m+2)-\frac{1}{2}\psi_{0}^{2}(n)+\nonumber\\
&&\psi_{0}(n-m+2)(\psi_{0}(n)-\psi_{0}(m)+\psi_{0}(1))+\left(\frac{4m-2}{m+n}+\frac{1}{n-m+1}+\frac{1}{n-m}+\frac{1}{n}-2\right)(\psi_{0}(n-m+2)-\nonumber\\
&&\psi_{0}(n))+\frac{(2n-2m+1)(\psi_{0}(m)-\psi_{0}(1))}{(n-m)(n-m+1)}-\frac{\psi_{0}(n)}{m}+\frac{5n-3mn-2m+2}{(m+n)n}-\frac{2n-m}{n(n-m)(n-m+1)}\bigg).
\end{eqnarray}
\end{widetext}
Similarly, the second sum in~(\ref{eq:IAe}) is simplified as
\begin{widetext}
\begin{eqnarray}\label{eq:IAs2}
&&\sum_{k=1}^{m-3}\frac{(n-1-k)!}{(m-3-k)!}\frac{1}{k(k+1)(k+2)(k+3)}=\sum_{k=1}^{m-3}\frac{(n-1-k)!}{(m-3-k)!2}\left(\frac{1}{3k}-\frac{1}{k+1}+\frac{1}{k+2}-\frac{1}{3(k+3)}\right)\nonumber\\
&\stackrel{(\ref{eq:sum11})}{=}&\frac{(n-1)!}{2m!}\left(\frac{m(m-1)(m-2)-n(n+1)(n+2)}{3}+mn(n-m+2)\right)\left(\psi_{0}(n)-\psi_{0}(n-m+2)-\frac{1}{n-m+2}\right)+\nonumber\\
&&\frac{(n-1)!}{36m!}(m-3)\left(11m^2+6n^2-15mn-18m+12n+4\right).
\end{eqnarray}
\end{widetext}
Inserting~(\ref{eq:IAs1}) and~(\ref{eq:IAs2}) into~(\ref{eq:IAe}), $I_{A}$ is simplified to
\begin{widetext}
\begin{eqnarray}\label{eq:IAf}
I_{A}&=&2mn(m+n)\sum_{k=1}^{m}\frac{\psi_{0}(n-m+k)}{k}+mn(m+n)\bigg(\psi_{1}(n-m+2)-\psi_{0}^{2}(n-m+2)+2\psi_{0}(n-m+2)(\psi_{0}(n)-\nonumber\\
&&\psi_{0}(m)+\psi_{0}(1))+\frac{2(2n-2m+1)\left(\psi_{0}(m)-\psi_{0}(1)\right)}{(n-m)(n-m+1)}\bigg)+\frac{a_{1}\psi_{0}(n)+a_{2}\psi_{0}(n-m+2)+a_{3}}{(n-m)(n-m+1)},
\end{eqnarray}
\begin{eqnarray}
a_{1}&=&\frac{n}{3}\left(9m^{3}n+9m^{3}-17m^{2}n^{2}-6m^{2}n-m^{2}+7mn^{3}-mn^{2}-10mn-2m+n^{4}-2n^{3}-n^{2}+2n\right),\label{eq:a1}\\
a_{2}&=&\frac{1}{3}\big(m^5 + 7 m^4 n + 2 m^4 - 26 m^3 n^2 - 26 m^3 n - m^3 + 26 m^2 n^3 + 18 m^2 n^2 + 3 m^2 n - 2 m^2 - 7 m n^4 + \nonumber\\
&&10 m n^3 + 15 m n^2 + 4 m n - n^5 - 4 n^4 - 5 n^3 - 2 n^2\big),\label{eq:a2}\\
a_{3}&=&\frac{1}{18}\big(\!-5 m^5 - 65 m^4 n + 14 m^4 + 139 m^3 n^2 + 169 m^3 n + 41 m^3 - 63 m^2 n^3 - 282 m^2 n^2 - 142 m^2 n - \nonumber\\
&&2 m^2 - 6 m n^4 + 87 m n^3 - 13 m n^2 - 40 m n - 12 m + 12 n^4 + 42 n^3 + 42 n^2 + 12 n\big).\label{eq:a3}
\end{eqnarray}
\end{widetext}
We now simplify $I_{B}$ in~(\ref{eq:IBe}), where the first two sums are
\begin{widetext}
\begin{eqnarray}\label{eq:IBs1}
&&\sum_{k=0}^{m-1}\left(\left(n-m+1+2k\right)\psi_{0}(n-m+1+k)+2k+1\right)^{2}+\sum_{k=0}^{m-2}\frac{2(k+1)}{n-m+1+k}((n-m+1+k)\times\nonumber\\
&&\psi_{0}(n-m+1+k)+n-m+2+3k/2)^{2}\nonumber\\
&=&6\sum_{k=0}^{m-2}k^{2}\psi_{0}^{2}(n-m+1+k)+2(3n-3m+4)\sum_{k=0}^{m-2}k\psi_{0}^{2}(n-m+1+k)+(n-m+1)(n-m+3)\times\nonumber\\
&&\sum_{k=0}^{m-2}\psi_{0}^{2}(n-m+1+k)+14\sum_{k=0}^{m-2}k^{2}\psi_{0}(n-m+1+k)+2(4n-4m+11)\sum_{k=0}^{m-2}k\psi_{0}(n-m+1+k)+\nonumber\\
&&2(3n-3m+5)\sum_{k=0}^{m-2}\psi_{0}(n-m+1+k)+\sum_{k=0}^{m-2}\frac{2(k+1)(n-m+2+3k/2)^{2}}{n-m+1+k}+\sum_{k=0}^{m-2}(2k+1)^{2}+\nonumber\\
&&\left((n+m-1)\psi_{0}(n)+2m-1\right)^{2}\nonumber\\
&\stackrel{(\ref{eq:sum1})-(\ref{eq:sum6})}{=}&mn(m+n-1)\psi_{0}^{2}(n)+\frac{1}{6}\left(3m^{3}+15m^{2}n+3mn^{2}-6mn-3m-n^{3}+n\right)\psi_{0}(n)+\frac{1}{6}(n-m-1)\times\nonumber\\
&&(n-m)(n-m+1)\psi_{0}(n-m+2)+\frac{1}{36}\left(35m^{3}+21m^{2}n+6mn^{2}-9mn-17m-12n^{2}+6n+6\right).
\end{eqnarray}
\end{widetext}
The remaining double sums in $I_{B}$ needs some preprocessing before the sum of the types in the appendix appear. Specifically, by shifting the inner sum $k\to k-j$, changing the summation order, and using partial fraction decomposition, we have
\begin{widetext}
\begin{eqnarray}\label{eq:IBs2}
&&\sum_{j=2}^{m-1}\sum_{k=0}^{m-j-1}\frac{2(n-m+k)!(k+j)!}{(n-m+k+j)!k!j^{2}}\left(\frac{n-m+1+k}{j-1}-\frac{k}{j+1}\right)^{2}\nonumber\\
&=&\sum_{k=2}^{m-1}\sum_{j=2}^{k}\frac{2(n-m+k-j)!k!}{(n-m+k)!(k-j)!j^{2}}\left(\frac{n-m+1+k-j}{j-1}-\frac{k-j}{j+1}\right)^{2}\nonumber\\
&=&\sum_{k=2}^{m-1}\sum_{j=2}^{k}\frac{2(n-m+k-j)!k!}{(n-m+k)!(k-j)!}\bigg((n-m+1+k-j)^{2}\left(\frac{1}{(j-1)^{2}}+\frac{1}{j^{2}}+\frac{2}{j}-\frac{2}{j-1}\right)+(k-j)^{2}\bigg(\frac{1}{(j+1)^{2}}+\frac{1}{j^{2}}+\nonumber\\
&&\frac{2}{j+1}-\frac{2}{j}\bigg)-(n-m+1+k-j)(k-j)\left(\frac{1}{j-1}-\frac{1}{j+1}-\frac{2}{j^{2}}\right)\!\bigg)=\mathcal{I}_{1}+\mathcal{I}_{2},
\end{eqnarray}
\end{widetext}
where $\mathcal{I}_{1}$ and $\mathcal{I}_{2}$ collect terms involving $1/j$ and $1/j^{2}$, respectively, as (the terms involving $1/j^{0}$ cancel)
\begin{widetext}
\begin{eqnarray}
\mathcal{I}_{1}&=&\sum_{k=2}^{m-1}\frac{2k!}{(n-m+k)!}\Bigg(\sum_{j=3}^{k+1}\frac{(n-m+2+k-j)!}{(k-j)!j}-\sum_{j=1}^{k-1}\frac{(n-m+k-j)!}{(k-2-j)!j}+(2k+1)\Bigg(\sum_{j=3}^{k+1}\frac{(n-m+1+k-j)!}{(k-j)!j}-\nonumber\\
&&\sum_{j=2}^{k}\frac{(n-m+k-j)!}{(k-1-j)!j}\Bigg)+2(n-m+1/2+k)\Bigg(\sum_{j=2}^{k}\frac{(n-m+1+k-j)!}{(k-j)!j}-\sum_{j=1}^{k-1}\frac{(n-m+k-j)!}{(k-1-j)!j}\Bigg)\!\Bigg),\\
\mathcal{I}_{2}&=&\sum_{k=2}^{m-1}\frac{2k!}{(n-m+k)!}\Bigg(2\sum_{j=2}^{k}\frac{(n-m+1+k-j)!}{(k-1-j)!j^{2}}+k\sum_{j=2}^{k}\frac{(n-m+k-j)!}{(k-1-j)!j^{2}}+(k+1)\sum_{j=3}^{k+1}\frac{(n-m+1+k-j)!}{(k-j)!j^{2}}+\nonumber\\
&&(n-m+k)\sum_{j=1}^{k-1}\frac{(n-m+k-j)!}{(k-1-j)!j^{2}}+(n-m+1+k)\sum_{j=2}^{k}\frac{(n-m+1+k-j)!}{(k-j)!j^{2}}\Bigg).
\end{eqnarray}
\end{widetext}
The sums in $\mathcal{I}_{1}$ are further simplified as
\begin{widetext}
\begin{eqnarray}\label{eq:IBs2i1}
\mathcal{I}_{1}&=&\sum_{k=2}^{m-1}\frac{2(3n-3m+4)(n-m+1+2k)k!}{(n-m+k)!}\sum_{j=2}^{k}\frac{(n-m+k-j)!}{(k-j)!j}-\sum_{k=2}^{m-1}\frac{4(n-m+2)k!}{(n-m+k)!}\times\nonumber\\
&&\sum_{j=2}^{k}\frac{(n-m+k-j)!}{(k-j)!}-\sum_{k=2}^{m-1}\frac{k(k-1)(5n-5m-1+9k)}{n-m+k}\nonumber\\
&\stackrel{(\ref{eq:sum10})}{=}&\sum_{k=2}^{m-1}\frac{2(3n-3m+4)(n-m+1+2k)k!}{(n-m+k)!}\sum_{j=2}^{k}\frac{(n-m+k-j)!}{(k-j)!j}+(n-m)\big(4m^{2}-8mn-m+4n^{2}+\nonumber\\
&&n-7\big)(\psi_{0}(n)-\psi_{0}(n-m+2))+\frac{m-2}{2(n-m+1)}\big(18m^{3}-38m^{2}n-19m^{2}+28mn^{2}+19mn-14m-\nonumber\\
&&8n^{3}-6n^{2}+13n+7\big)\nonumber\\
&\stackrel{(\ref{eq:sum11})}{=}&2(3n-3m+4)\left(\sum_{k=2}^{m-1}(n-m+1+2k)\psi_{0}(n-m+1+k)-(mn+2m-2n-4)\psi_{0}(n-m+2)\right)-\nonumber\\
&&(2n-2m-1)(n-m)(n-m+1)(\psi_{0}(n)-\psi_{0}(n-m+2))+\frac{6m^{2}-16mn-7m+4n^{2}+6n+7}{2(m-2)^{-1}}\nonumber\\
&\stackrel{(\ref{eq:sum1})-(\ref{eq:sum2})}{=}&\left(2m^{3}-12m^{2}n-m^{2}+12mn^{2}+10mn-m-2n^{3}-n^{2}+n\right)(\psi_{0}(n)-\psi_{0}(n-m+2))+\nonumber\\
&&\frac{1}{2}(m-2)\left(12m^{2}-22mn-9m+4n^{2}-1\right).
\end{eqnarray}
\end{widetext}
The sums in $\mathcal{I}_{2}$ are further simplified as
\begin{widetext}
\begin{eqnarray}\label{eq:IBs2i2}
\mathcal{I}_{2}&=&\sum_{k=2}^{m-1}\frac{2\left(m^{2}-2mn-3m+n^{2}+3n+2-6(m-n-1)k+6k^{2}\right)k!}{(n-m+k)!}\sum_{j=2}^{k}\frac{(n-m+k-j)!}{(k-j)!j^{2}}-\nonumber\\
&&\sum_{k=2}^{m-1}\frac{8(n-m+1+2k)k!}{(n-m+k)!}\sum_{j=2}^{k}\frac{(n-m+k-j)!}{(k-j)!j}+\sum_{k=2}^{m-1}\frac{4k!}{(n-m+k)!}\sum_{j=2}^{k}\frac{(n-m+k-j)!}{(k-j)!}+\nonumber\\
&&\sum_{k=2}^{m-1}\frac{k(k-1)(4n-4m-1+3k)}{2(n-m+k)}\nonumber\\
&\stackrel{(\ref{eq:sum10})-(\ref{eq:sum11})}{=}&\sum_{k=2}^{m-1}\frac{2\left(m^{2}-2mn-3m+n^{2}+3n+2-6(m-n-1)k+6k^{2}\right)k!}{(n-m+k)!}\sum_{j=2}^{k}\frac{(n-m+k-j)!}{(k-j)!j^{2}}-\nonumber\\
&&\frac{1}{2}\left(m^{3}-3m^{2}n-16m^{2}+3mn^{2}+48mn-9m-n^{3}-16n^{2}+9n\right)(\psi_{0}(n)-\psi_{0}(n-m+2))+\nonumber\\
&&\frac{m-2}{4(n-m+1)}\left(m^{3}-6m^{2}n-77m^{2}+7mn^{2}+112mn+19m-2n^{3}-33n^{2}+2n+9\right)\nonumber\\
&\stackrel{(\ref{eq:sum12})}{=}&\sum_{k=2}^{m-1}\left((n-m+2)(n-m+1)+6(n-m+1)k+6k^{2}\right)\Bigg(\sum_{j=1}^{k}\frac{2\psi_{0}(n-m+j)}{j}-\psi_{1}(n-m+1+k)-\nonumber\\
&&\psi_{0}^{2}(n-m+1+k)+\psi_{1}(n-m+1)+\psi_{0}^{2}(n-m+1)+2\psi_{0}(n-m)(\psi_{0}(n-m+1+k)-\psi_{0}(k+1)-\nonumber\\
&&\psi_{0}(n-m+1)+\psi_{0}(1))\!\Bigg)-\frac{1}{2}\left(5m^{3}-15m^{2}n-4m^{2}+15mn^{2}+24mn-m-5n^{3}-4n^{2}+n\right)(\psi_{0}(n)-\nonumber\\
&&\psi_{0}(n-m+2))+\frac{(m-2)\left(25m^{3}-46m^{2}n-45m^{2}+31mn^{2}+48mn+27m-10n^{3}-17n^{2}-38n-55\right)}{4(n-m+1)}\nonumber\\
&\stackrel{(\ref{eq:sum1})-(\ref{eq:sum9})}{=}&2mn(m+n)\sum_{k=1}^{m}\frac{\psi_{0}(n-m+k)}{k}+mn(m+n)\bigg(\!-\psi_{1}(n)+\psi_{1}(n-m+2)-\psi_{0}^{2}(n)-\psi_{0}^{2}(n-m+2)+\nonumber\\
&&2\psi_{0}(n-m+2)(\psi_{0}(n)-\psi_{0}(m)+\psi_{0}(1))+\frac{2(2n-2m+1)\left(\psi_{0}(m)-\psi_{0}(1)\right)}{(n-m)(n-m+1)}\bigg)+\nonumber\\
&&\frac{b_{1}\psi_{0}(n)+b_{2}\psi_{0}(n-m+2)+b_{3}}{(n-m)(n-m+1)},
\end{eqnarray}
\end{widetext}
where we also changed the summation order between $j$ and $k$ to arrive at the last equality, and $b_{1}$, $b_{2}$, $b_{3}$ are
\begin{widetext}
\begin{eqnarray}
b_{1}&=&\frac{1}{2}\big(\!-5m^{5}+29m^{4}n+5m^{4}-62m^{3}n^{2}-40m^{3}n+m^{3}+62m^{2}n^{3}+86m^{2}n^{2}+17m^{2}n-m^{2}-29mn^{4}-\nonumber\\
&&56mn^{3}-25mn^{2}+2mn+5n^{5}+5n^{4}-n^{3}-n^{2}\big),\\
b_{2}&=&\frac{1}{2}\big(5m^{5}-29m^{4}n-5m^{4}+62m^{3}n^{2}+36m^{3}n-m^{3}-62m^{2}n^{3}-82m^{2}n^{2}-13m^{2}n+m^{2}+29mn^{4}+\nonumber\\
&&60mn^{3}+25mn^{2}-2mn-5n^{5}-9n^{4}-3n^{3}+n^{2}\big),\\
b_{3}&=&\frac{1}{4}\big(\!-29m^{5}+83m^{4}n+95m^{4}-89m^{3}n^{2}-247m^{3}n-89m^{3}+45m^{2}n^{3}+243m^{2}n^{2}+187m^{2}n+29m^{2}-\nonumber\\
&&10mn^{4}-111mn^{3}-140mn^{2}-33mn+2m+20n^{4}+26n^{3}+4n^{2}-2n\big).
\end{eqnarray}
\end{widetext}
With $\mathcal{I}_{1}$ and $\mathcal{I}_{2}$ being simplified as in~(\ref{eq:IBs2i1}) and~(\ref{eq:IBs2i2}), respectively, we now insert~(\ref{eq:IBs1}) and~(\ref{eq:IBs2}) into~(\ref{eq:IBe}) to obtain
\begin{widetext}
\begin{eqnarray}\label{eq:IBf}
I_{B}&=&2mn(m+n)\sum_{k=1}^{m}\frac{\psi_{0}(n-m+k)}{k}+mn(m+n)\bigg(\!-\psi_{1}(n)+\psi_{1}(n-m+2)-\frac{\psi_{0}^{2}(n)}{m+n}-\psi_{0}^{2}(n-m+2)+\nonumber\\
&&2\psi_{0}(n-m+2)(\psi_{0}(n)-\psi_{0}(m)+\psi_{0}(1))+\frac{2(2n-2m+1)\left(\psi_{0}(m)-\psi_{0}(1)\right)}{(n-m)(n-m+1)}\bigg)+\nonumber\\
&&\frac{b_{4}\psi_{0}(n)+b_{5}\psi_{0}(n-m+2)+b_{6}}{(n-m)(n-m+1)},
\end{eqnarray}
\begin{eqnarray}
b_{4}&=&\frac{1}{3}\big(\!-3m^{4}+9m^{3}n^{2}+9m^{3}n-17m^{2}n^{3}+3m^{2}n^{2}+8m^{2}n+3m^{2}+7mn^{4}-7mn^{3}-19mn^{2}-5mn+\nonumber\\
&&n^{5}-2n^{4}-n^{3}+2n^{2}\big),\label{eq:b4}\\
b_{5}&=&\frac{1}{3}\big(m^{5}+7m^{4}n+2m^{4}-26m^{3}n^{2}-26m^{3}n-m^{3}+26m^{2}n^{3}+18m^{2}n^{2}+3m^{2}n-2m^{2}-7mn^{4}+\nonumber\\
&&10mn^{3}+15mn^{2}+4mn-n^{5}-4n^{4}-5n^{3}-2n^{2}\big),\label{eq:b5}\\
b_{6}&=&\frac{1}{18}\big(\!-5m^{5}-65m^{4}n+5m^{4}+139m^{3}n^{2}+187m^{3}n+41m^{3}-63m^{2}n^{3}-291m^{2}n^{2}-133m^{2}n+7m^{2}-\nonumber\\
&&6mn^{4}+87mn^{3}-22mn^{2}-49mn-12m+12n^{4}+42n^{3}+42n^{2}+12n\big).\label{eq:b6}
\end{eqnarray}
\end{widetext}
We observe that $I_{A}$ in~(\ref{eq:IAf}) and $I_{B}$ in~(\ref{eq:IBf}) share many common terms, where by inserting~(\ref{eq:IAf}) and~(\ref{eq:IBf}) into~(\ref{eq:i2m1}) the remaining terms of the induced variance $\mathbb{E}_{g}\!\left[T^{2}\right]$ are
\begin{widetext}
\begin{eqnarray*}
\mathbb{E}_{g}\!\left[T^{2}\right]&=&mn(m+n)\psi_{1}(n)+mn(mn+1)\psi_{0}^{2}(n)+m^{2}(m+1)n\psi_{0}(n)+\frac{1}{4}m^{2}(m+1)^{2}+\\
&&\frac{(a_{1}-b_{4})\psi_{0}(n)+(a_{2}-b_{5})\psi_{0}(n-m+2)+a_{3}-b_{6}}{(n-m)(n-m+1)}\\
&=&mn(m+n)\psi_{1}(n)+mn(mn+1)\psi_{0}^{2}(n)+m\left(m^{2}n+mn+m+2n+1\right)\psi_{0}(n)+\frac{1}{4}m(m+1)\left(m^2+m+2\right),
\end{eqnarray*}
\end{widetext}
where we have used the results
\begin{eqnarray*}
a_{1}-b_{4}&=&m(n-m)(n-m+1)(2n+m+1),\\
a_{2}-b_{5}&=&0,\\
a_{3}-b_{6}&=&\frac{1}{2}m(m+1)(n-m)(n-m+1),
\end{eqnarray*}
obtained by comparing~(\ref{eq:a1})--(\ref{eq:a3}) to~(\ref{eq:b4})--(\ref{eq:b6}). This completes the proof of the induced conjecture~(\ref{eq:eqc}) in the case $n>m\geq3$ and hence the VPO's conjecture~(\ref{eq:vNv}) for the same case.

\begin{table*}[t!]
\caption{Special Cases}\centering
\ra{2.0}
\begin{ruledtabular}
\begin{tabular}{l|c|l}
\multirow{3}{*}{$m=1$} & $I_{A}$ & $n(n+1)\psi_{1}(n)+n(n+1)\psi_{0}^{2}(n)+(4n+2)\psi_{0}(n)+2$ \\
                       & $I_{B}$ & $\left(n\psi_{0}(n)+1\right)^{2}$ \\
                       & $\mathbb{E}_{g}\!\left[T^{2}\right]$ & $n(n+1)\psi_{1}(n)+n(n+1)\psi_{0}^{2}(n)+(4n+2)\psi_{0}(n)+2$ \\ \hline
\multirow{3}{*}{$m=2$} & $I_{A}$ & $2\left(n(n+2)\psi_{1}(n)+n(n+2)\psi_{0}^{2}(n)+(7n+4)\psi_{0}(n)+n+5\right)$ \\
                       & $I_{B}$ & $2n(n+1)\psi_{0}^{2}(n)+2(5n+1)\psi_{0}(n)+2n+7$ \\
                       & $\mathbb{E}_{g}\!\left[T^{2}\right]$ & $2\left(n(n+2)\psi_{1}(n)+n(2n+1)\psi_{0}^{2}(n)+(8n+3)\psi_{0}(n)+6\right)$ \\ \hline
\multirow{3}{*}{$m=n$}&$I_{A}$&$\frac{1}{9}\left(-18n^{3}\psi_{1}(n)+36n^{3}\psi_{1}(1)+18n^{3}\psi_{0}^{2}(n)+6n\left(5n^{2}+3n+1\right)\psi_{0}(n)-43n^{3}+33n^{2}+22n+6\right)$\\
                      &$I_{B}$&$\frac{1}{18}\left(-72n^{3}\psi_{1}(n)+72n^{3}\psi_{1}(1)+18(2n-1)n^{2}\psi_{0}^{2}(n)+6n\left(10n^{2}-3n-1\right)\psi_{0}(n)-86n^{3}+57n^{2}+35n+12\right)$\\
                      &$\mathbb{E}_{g}\!\left[T^{2}\right]$&$\frac{1}{4}\left(8n^{3}\psi_{1}(n)+4n^{2}\left(n^{2}+1\right)\psi_{0}^{2}(n)+4n\left(n^{3}+n^{2}+3n+1\right)\psi_{0}(n)+n(n+1)\left(n^{2}+n+2\right)\right)$\\
\end{tabular}\label{t:sc}
\end{ruledtabular}
\end{table*}

Since $m\leq n$, the remaining cases to be shown are $m=1$, $m=2$, and $m=n$, where $I_{A}$ in~(\ref{eq:IAe}) and $I_{B}$ in~(\ref{eq:IBe}) can be directly computed. We list the simplified expressions for $I_{A}$, $I_{B}$, and the induced variance $\mathbb{E}_{g}\!\left[T^{2}\right]$ in Table~\ref{t:sc} as shown on top of the next page. Each of the special cases is proven by comparing the expression of $\mathbb{E}_{g}\!\left[T^{2}\right]$ in Table~\ref{t:sc} with that of the corresponding induced conjecture~(\ref{eq:eqc}). We complete the proof of the VPO's conjecture~(\ref{eq:vNv}).

\begin{acknowledgments}
The author wishes to thank Michael Milgram, Gregory Schehr, and Yu Xiang for the inspiring discussion.
\end{acknowledgments}

\appendix*

\onecolumngrid
\begin{widetext}
\section{Finite Sum Identities Useful in Section~\ref{sec:clear}}
\begin{equation}
\sum_{k=1}^{n}\psi_{0}(k+l)=(n+l)\psi_{0}(n+l)-l\psi_{0}(l)-n.\label{eq:sum1}
\end{equation}
\hrulefill
\begin{equation}
\sum_{k=1}^{n}k\psi_{0}(k+l)=\frac{1}{2}\left(n^{2}+n-l^{2}+l\right)\psi_{0}(n+l)+\frac{1}{2}l(l-1)\psi_{0}(l)+\frac{1}{4}n(-n+2l-1).\label{eq:sum2}
\end{equation}
\hrulefill
\begin{eqnarray}
\sum_{k=1}^{n}k^{2}\psi_{0}(k+l)&=&\frac{1}{6}\left(2n^{3}+3n^{2}+n+2l^{3}-3l^{2}+l\right)\psi_{0}(n+l)-\frac{1}{6}l\left(2l^{2}-3l+1\right)\psi_{0}(l)+\nonumber\\
&&\frac{1}{36}n\left(-4n^{2}+6nl-3n-12l^{2}+12l+1\right).\label{eq:sum3}
\end{eqnarray}
\hrulefill
\begin{equation}
\sum_{k=1}^{n}\psi_{0}^{2}(k+l)=\left(n+l\right)\psi_{0}^{2}(n+l)-\left(2n+2l-1\right)\psi_{0}(n+l)-l\psi_{0}^{2}(l)+(2l-1)\psi_{0}(l)+2n.\label{eq:sum4}
\end{equation}
\hrulefill
\begin{eqnarray}
\sum_{k=1}^{n}k\psi_{0}^{2}(k+l)&=&\frac{1}{2}\left(n^{2}+n-l^{2}+l\right)\psi_{0}^{2}(n+l)+\frac{1}{2}\left(-n^{2}+2nl-n+3l^{2}-3l+1\right)\psi_{0}(n+l)+\nonumber\\
&&\frac{1}{2}l(l-1)\psi_{0}^{2}(l)-\frac{1}{2}\left(3l^{2}-3l+1\right)\psi_{0}(l)+\frac{1}{4}n(n-6l+3).\label{eq:sum5}
\end{eqnarray}
\hrulefill
\begin{eqnarray}
\sum_{k=1}^{n}k^{2}\psi_{0}^{2}(k+l)&=&\frac{1}{6}\left(2n^{3}+3n^{2}+n+2l^{3}-3l^{2}+l\right)\psi_{0}^{2}(n+l)-\frac{1}{18}\big(4n^{3}-6n^{2}l+3n^{2}+12nl^{2}-\nonumber\\
&&-12nl-n+22l^{3}-33l^{2}+17l-3\big)\psi_{0}(n+l)-\frac{1}{6}l\left(2l^{2}-3l+1\right)\psi_{0}^{2}(l)+\nonumber\\
&&\frac{1}{18}\big(22l^{3}-33l^{2}+17l-3\big)\psi_{0}(l)+\frac{1}{108}n\left(8n^{2}-30nl+15n+132l^{2}-132l+25\right).\label{eq:sum6}
\end{eqnarray}
\hrulefill
\begin{equation}
\sum_{k=1}^{n}\psi_{1}(k+l)=(n+l)\psi_{1}(n+l)-l\psi_{1}(l)+\psi_{0}(n+l)-\psi_{0}(l).\label{eq:sum7}
\end{equation}
\hrulefill
\begin{eqnarray}
\sum_{k=1}^{n}k\psi_{1}(k+l)&=&\frac{1}{2}\left(n^{2}+n-l^{2}+l\right)\psi_{1}(n+l)+\frac{1}{2}l(l-1)\psi_{1}(l)-\frac{1}{2}(2l-1)\psi_{0}(n+l)+\nonumber\\
&&\frac{1}{2}(2l-1)\psi_{0}(l)+\frac{1}{2}n.\label{eq:sum8}
\end{eqnarray}
\hrulefill
\begin{eqnarray}
\sum_{k=1}^{n}k^{2}\psi_{1}(k+l)&=&\frac{1}{6}\left(2n^{3}+3n^{2}+n+2l^{3}-3l^{2}+l\right)\psi_{1}(n+l)-\frac{1}{6}l\left(2l^{2}-3l+1\right)\psi_{1}(l)+\nonumber\\
&&\frac{1}{6}\left(6l^{2}-6l+1\right)\psi_{0}(n+l)-\frac{1}{6}\left(6l^{2}-6l+1\right)\psi_{0}(l)+\frac{1}{6}n(n-4l+2).\label{eq:sum9}
\end{eqnarray}
\hrulefill
\begin{equation}
\sum_{k=1}^{m}\frac{(n-k)!}{(m-k)!}=\frac{(n-1)!}{(m-1)!}\frac{n}{n-m+1}.\label{eq:sum10}
\end{equation}
\hrulefill
\begin{equation}
\sum_{k=1}^{m}\frac{(n-k)!}{(m-k)!}\frac{1}{k}=\frac{n!}{m!}\left(\psi_{0}(n+1)-\psi_{0}(n-m+1)\right).\label{eq:sum11}
\end{equation}
\hrulefill
\begin{eqnarray}
\sum_{k=1}^{m}\frac{(n-k)!}{(m-k)!}\frac{1}{k^2}&=&\frac{n!}{m!}\sum_{k=1}^{m}\frac{\psi_{0}(n-m+k)}{k}+\frac{n!}{2m!}\big(\psi_{1}(n-m+1)-\psi_{1}(n+1)+\psi_{0}^{2}(n-m+1)-\nonumber\\
&&\psi_{0}^{2}(n+1)\big)+\frac{n!}{m!}\psi_{0}(n-m)\left(\psi_{0}(n+1)-\psi_{0}(m+1)-\psi_{0}(n-m+1)+\psi_{0}(1)\right).\label{eq:sum12}
\end{eqnarray}
\end{widetext}

\subsubsection*{Some Remarks on the Identities in the Appendix}
The formulas of finite sums of polygamma functions of the types~(\ref{eq:sum1})--(\ref{eq:sum9}) are straightforward to show. The proofs essentially involve changing the order of the sums and making use of the lower order sums already obtained in a recursive manner. In particular, the formulas~(\ref{eq:sum1})--(\ref{eq:sum4}) are available in~\cite[ch.~5.1]{Brychkov}. The formulas~(\ref{eq:sum5})--(\ref{eq:sum9}) can be read off from the expressions in~\cite[p.~861]{Spiess1990} by keeping in mind the difference between polygamma functions~(\ref{eq:dg}),~(\ref{eq:tg}) and harmonic numbers.

The last three formulas~(\ref{eq:sum10})--(\ref{eq:sum12}) play a crucial role in the simplification in Sec.~\ref{sec:clear} as they connect some of the sums in~(\ref{eq:IAe}) and~(\ref{eq:IBe}) to polygamma functions. The first of them~(\ref{eq:sum10}) is known as Chu-Vandermonde identity~\cite[p.~99]{Luke}. The next formula~(\ref{eq:sum11}) can be established as follows. First, the identity~(\ref{eq:dgre}) implies that
\begin{equation}\label{eq:wy1}
\sum_{k=1}^{m}\frac{(n-k)!}{(m-k)!}\frac{1}{k}=\sum_{k=1}^{m}\frac{(n-k)!}{(m-k)!}\left(\psi_{0}(k+1)-\psi_{0}(k)\right).
\end{equation}
By using the definition of digamma function~(\ref{eq:dg}), changing the order of sums, and evoking Chu-Vandermonde identity~(\ref{eq:sum10}), the first term in~(\ref{eq:wy1}) is represented as
\begin{eqnarray}\label{eq:wyi}
&&\sum_{k=1}^{m}\frac{(n-k)!}{(m-k)!}\psi_{0}(k+1)=\frac{n+1}{n-m+1}\sum_{k=1}^{m}\frac{(n-k)!}{(m-k)!}\frac{1}{k}-\nonumber\\
&&\frac{n!}{(n-m+1)(m-1)!}\left(\gamma+\frac{1}{n-m+1}\right).
\end{eqnarray}
Similarly, we have
\begin{eqnarray}\label{eq:wyi1}
&&\sum_{k=1}^{m}\frac{(n-k)!}{(m-k)!}\psi_{0}(k)=\frac{n}{n-m+1}\sum_{k=1}^{m-1}\frac{(n-1-k)!}{(m-1-k)!}\frac{1}{k}-\nonumber\\
&&\frac{(n-1)!}{(n-m+1)(m-1)!}\left(\gamma n+\frac{m-1}{n-m+1}\right).
\end{eqnarray}
Inserting~(\ref{eq:wyi}) and~(\ref{eq:wyi1}) into~(\ref{eq:wy1}), we obtain a recurrence relation of the sum~(\ref{eq:sum11}) as
\begin{equation}\label{eq:wy1r}
s(m,n)=\frac{(n-1)!}{m!}+\frac{n}{m}s(m-1,n-1),
\end{equation}
where we denote
\begin{equation}
s(m,n)=\sum_{k=1}^{m}\frac{(n-k)!}{(m-k)!}\frac{1}{k}.
\end{equation}
Finally, by iterating $m-1$ times the relation~(\ref{eq:wy1r}), we arrive at
\begin{eqnarray}
s(m,n)&=&\frac{n!}{m!}\left(\frac{1}{n}+\frac{1}{n-1}+\dots+\frac{1}{n-m+1}\right)+\nonumber\\
&&\frac{n(n-1)\cdots(n-m+1)}{m(m-1)\cdots1}s(0,n-m)\nonumber\\
&=&\frac{n!}{m!}\left(\psi_{0}(n+1)-\psi_{0}(n-m+1)\right),
\end{eqnarray}
where we have used the fact that $s(0,n-m)=0$. Note that the formula~(\ref{eq:sum11}) can be also obtained via its connection to a hypergeometric function of unit argument as~\cite[p.~111]{Luke}
\begin{eqnarray*}
\sum_{k=1}^{m}\frac{(n-k)!}{(m-k)!}\frac{1}{k}&=&\frac{(n-1)!}{(m-1)!}~\!_{3}F_{2}\left(1,1,1-m;2,1-n;1\right)\\
&=&\frac{n!}{m!}\left(\psi_{0}(n+1)-\psi_{0}(n-m+1)\right).
\end{eqnarray*}
To prove the last formula~(\ref{eq:sum12}), we first observe from~(\ref{eq:tgre}) that
\begin{equation}\label{eq:wy2}
\sum_{k=1}^{m}\frac{(n-k)!}{(m-k)!}\frac{1}{k^2}=\sum_{k=1}^{m}\frac{(n-k)!}{(m-k)!}\left(\psi_{1}(k)-\psi_{1}(k+1)\right).
\end{equation}
Following the same idea that has led to~(\ref{eq:wy1r}), we also obtain a recurrence relation in this case as
\begin{eqnarray}\label{eq:wy2r}
t(m,n)&=&\frac{(n-1)!(n-m)}{m!m}(\psi_{0}(n)-\psi_{0}(n-m))+\nonumber\\
&&\frac{n}{m}t(m-1,n-1),
\end{eqnarray}
where we denote
\begin{equation}
t(m,n)=\sum_{k=1}^{m}\frac{(n-k)!}{(m-k)!}\frac{1}{k^2}.
\end{equation}
Iterating $m-1$ times the relation~(\ref{eq:wy2r}), we arrive at
\begin{eqnarray}
&&\sum_{k=1}^{m}\frac{(n-k)!}{(m-k)!}\frac{1}{k^2}=\frac{n!}{m!}\sum_{k=1}^{m}\frac{\psi_{0}(n-m+k)}{k}-\nonumber\\
&&\frac{n!}{m!}\sum_{k=1}^{m}\frac{\psi_{0}(n-m+k)}{n-m+k}+\frac{n!}{m!}\psi_{0}(n-m)(\psi_{0}(n+1)-\nonumber\\
&&\psi_{0}(m+1)-\psi_{0}(n-m+1)+\psi_{0}(1)),
\end{eqnarray}
where by using the identity~\cite[eq.~(23)]{Milgram2004}
\begin{eqnarray}
&&\sum_{k=1}^{m}\frac{\psi(n-m+k)}{n-m+k}=\frac{1}{2}\big(\psi_{1}(n+1)-\psi_{1}(n-m+1)+\nonumber\\
&&\psi_{0}^{2}(n+1)-\psi_{0}^{2}(n-m+1)\big),
\end{eqnarray}
we obtain the claimed formula~(\ref{eq:sum12}). Though the expression~(\ref{eq:sum12}) still contains a sum of digamma functions that may not be further simplified, it is sufficient for the simplification purpose. As shown in Sec.~\ref{sec:clear}, the terms involving this remaining sum cancel each other. Finally, we note that as a result of the relation to the hypergeometric function
\begin{equation*}
\sum_{k=1}^{m}\frac{(n-k)!}{(m-k)!}\frac{1}{k^2}=\frac{(n-1)!}{(m-1)!}~\!_{4}F_{3}\left(1,1,1,1-m;2,2,1-n;1\right)
\end{equation*}
the formula~(\ref{eq:sum12}) implies a byproduct that generalizes a result of Luke~\cite[p.~111]{Luke} as
\begin{eqnarray}
&&_{4}F_{3}\left(1,1,1,1-m;2,2,1-n;1\right)=\nonumber\\
&&\frac{n}{m}\sum_{k=1}^{m}\frac{\psi_{0}(n-m+k)}{k}+\frac{n}{2m}\big(\psi_{1}(n-m+1)-\psi_{1}(n+1)+\nonumber\\
&&\psi_{0}^{2}(n-m+1)-\psi_{0}^{2}(n+1)\big)+\frac{n}{m}\psi_{0}(n-m)(\psi_{0}(n+1)-\nonumber\\
&&\psi_{0}(m+1)-\psi_{0}(n-m+1)+\psi_{0}(1)),
\end{eqnarray}
which may be of independent interest.

% \bibliography{apssamp}

\providecommand{\noopsort}[1]{}\providecommand{\singleletter}[1]{#1}%

\end{document}